\documentclass[journal]{IEEEtran}
\usepackage{times,amsmath}
\usepackage{amssymb}
\usepackage{steinmetz}
\usepackage{pstool}
\usepackage{verbatim}
\usepackage{booktabs}
\usepackage{subfigure}
\usepackage{subcaption}
\usepackage{subfigure}
\usepackage{multirow}
\usepackage{enumerate}
\usepackage{graphicx}
\usepackage{multirow}
\usepackage{subfig}
\usepackage{adjustbox} 
\usepackage[utf8]{inputenc}
\usepackage{tikz}
\usetikzlibrary{arrows, positioning, shapes, decorations.pathreplacing}

\usepackage{tabularx} % Add this in preamble

\usepackage{float}

\usepackage{mwe}
\usepackage{bigints}
\let\labelindent\relax
\usepackage{enumitem}
\usepackage{adjustbox}
\usepackage{booktabs}
\usepackage{pdflscape}

\usepackage{MnSymbol}
\usepackage{stfloats} 
\usepackage[table]{xcolor}
\usepackage[square, comma, sort&compress, numbers]{natbib}
\usepackage{nohyperref}
\usepackage{algorithm,algorithmic}
\usepackage{epstopdf}
\usepackage{mathtools, cuted}
\graphicspath{{Figures/3xmulti/}{Figures/3xbinary/}}
\usepackage{algorithm}
\usepackage{algorithmic}
\usepackage{xcolor} % Include xcolor package
\usepackage[table]{xcolor}
\definecolor{lightgray}{rgb}{0.9, 0.9, 0.9} % Define
\definecolor{lightgreen}{RGB}{180, 238, 180} 
\definecolor{green2}{RGB}{100, 238, 100} 
\definecolor{lightred}{RGB}{255, 182, 193}
\definecolor{lightyellow}{RGB}{255, 255, 100} 
\definecolor{lightlightyellow}{RGB}{255, 255, 220} 

\usepackage{rotating}
\usepackage{placeins}

\usepackage{tabularx}
\usepackage{array}
\usepackage{ragged2e}

\usepackage{tikz}
\usepackage{pgfplots}
\usepackage{xcolor}
\usetikzlibrary{arrows.meta, positioning, shapes, calc}
\pgfplotsset{compat=1.18}

\definecolor{open}{RGB}{60,150,60}
\definecolor{limited}{RGB}{240,180,60}
\definecolor{closed}{RGB}{200,60,60}

\begin{document}

\title{ECG Foundation Models and Medical LLMs for Agentic Cardiovascular Intelligence at the Edge: A Review and Outlook}

%\title{ECG Foundation Models and Medical LLMs-based Agentic AI for Cardiovascular Intelligence at the Edge: Review and Outlook}

%\title{Agentic AI for ECG Intelligence at the Edge: A Review of ECG Foundation Models and Medical LLMs}

%\title{Foundation Models and LLMs for ECG Intelligence at the Edge: A Detailed Review and Outlook}

%\title{Large Language Models for Analysis of Electrocardiogram Signals: A Detailed Review}

%\title{LLM-Driven ECG Intelligence at the Edge: A Detailed Review and Outlook}

\begin{comment}
From Signals to Reasoning: Large Language Models for Next-Generation ECG Intelligence

ECG Meets Foundation Models: A Translational Survey of LLM-Driven Cardiovascular AI

Beyond Arrhythmia Detection: LLM-Enabled ECG Intelligence for Scalable Cardiovascular Care

Reasoning Over Heartbeats: Large Language Models in ECG Analysis and Edge Cardiology

Foundation Models for the Failing Heart: A Survey of LLM-Driven ECG Intelligence

Edge-Aware Cardiovascular AI: Large Language Models for Real-World ECG Systems

Toward Continuous Cardiac Intelligence: LLMs, ECG Foundation Models, and Wearable AI

From Waveforms to Wisdom: A System-Level Survey of LLM-Enabled ECG Analysis
\end{comment}

\author{
\IEEEauthorblockN{
Mudassir Hasan Khan\IEEEauthorrefmark{1},
Ahmad Nayfeh\IEEEauthorrefmark{1},
Mudassir Masood\IEEEauthorrefmark{1}, {\it Senior member, IEEE}, 
Ali Ahmad Al-Shaikhi\IEEEauthorrefmark{1},
Muhammad Mahboob Ur Rahman\IEEEauthorrefmark{2}, {\it Senior member, IEEE},
Tareq Y. Al-Naffouri\IEEEauthorrefmark{2}, {\it Fellow, IEEE}
}

%\IEEEauthorblockA{\IEEEauthorrefmark{1} Computer, Electrical and Mathematical Sciences and Engineering Division (CEMSE), King Abdullah University of Science and Technology, Thuwal 23955, Saudi Arabia\\ 
%\IEEEauthorrefmark{1}\{muhammad.rahman,tareq.alnaffouri\}@kaust.edu.sa }
\thanks{
Mudassir Hasan Khan, Ahmad Nayfeh, Mudassir Masood, and Ali Ahmad Al-Shaikhi are affiliated with Electrical Engineering Department, King Fahd University of Petroleum \& Minerals (KFUPM), Dhahran 31261, Saudi Arabia. 
Ali Ahmad Al-Shaikhi is also affiliated with Interdisciplinary Research Center for Communication Systems and Sensing (IRC-CSS), KFUPM, Saudi Arabia.
%Email: shaikhi@kfupm.edu.sa
Muhammad Mahboob Ur Rahman and Tareq Y. Al-Naffouri are affiliated with Computer, Electrical and Mathematical Sciences and Engineering (CEMSE) Division, King Abdullah University of Science and Technology (KAUST), Thuwal 23955, Saudi Arabia. Email: \{muhammad.rahman\}@kaust.edu.sa. \\

The research reported in this publication was supported in part by: 
KFUPM, Saudi Arabia, the IRC-CSS at KFUPM, Saudi Arabia
and
funding from KAUST Center of Excellence for Smart Health (KCSH), under award number 5932. \\
}
}

\maketitle

\begin{abstract}

%v6 (for ieee trans. on consumer electronics)
Electrocardiogram (ECG) foundation models represent a paradigm shift from task-specific pipelines to generalizable architectures pre-trained on large-scale unlabeled waveform data. This survey presents a unified and deployment-aware review of foundation models and medical large language models (LLMs) for ECG intelligence in cardiovascular disease (CVD) diagnosis, monitoring, and clinical decision support. 
The central thesis of this survey paper is that next-generation cardiovascular AI systems will be inherently agentic, requiring the synergistic integration of two complementary model classes: (i) ECG foundation models that act as signal-level interpreters, learning rich electrophysiological representations via self-supervised and multimodal pretraining, and (ii) medical LLMs, trained on biomedical text corpora, that function as knowledge-based reasoning backbones for contextual inference, guideline alignment, and clinical decision support.
Thus, the survey systematically reviews existing pool of generalist medical LLMs, as well as ECG foundation models that utilize techniques such as self-supervised learning, multimodal ECG-language alignment, vision transformer architectures, and possess capabilities such as zero-shot classification, automated report generation, and longitudinal risk modeling. 
Recognizing the constraints of consumer-grade wearable edge devices, we further examine model optimization techniques such as quantization, pruning, knowledge distillation, as well as the role of small language models in enabling low-latency, energy-efficient, and privacy-preserving ECG intelligence on edge platforms such as smartwatches.
Finally, we outline future directions in multimodal ECG foundation models, agent-driven monitoring, and explainable, secure edge intelligence, with particular emphasis on real-time, on-device cardiovascular analytics in consumer electronics ecosystems.

\end{abstract}

\begin{IEEEkeywords}
cardiovascular diseases, ECG, foundation model, medical LLM, model optimization, edge AI, agentic AI.      
\end{IEEEkeywords}

\section{Introduction}

Cardiovascular diseases (CVDs) remain the leading cause of mortality worldwide, accounting for a substantial proportion of preventable deaths and imposing significant clinical and economic burdens across healthcare systems \cite{deaton2011global}. The electrocardiogram (ECG) is central to the screening, diagnosis, and longitudinal monitoring of CVDs, offering a non-invasive, cost-effective modality for capturing cardiac electrophysiological activity with high temporal resolution and established clinical interpretability \cite{de1998prognostic}. ECG analysis underpins the detection of arrhythmias, ischemia, conduction abnormalities, and autonomic dysfunction, and is increasingly embedded in digital health ecosystems ranging from hospital telemetry to wearable devices. However, its clinical utility remains constrained by dependence on expert interpretation, which is labor-intensive, subject to variability, and difficult to scale in high-throughput or resource-limited settings \cite{rafie2021ecg}. These limitations have driven sustained interest in automated ECG intelligence capable of delivering accurate, timely, and accessible cardiovascular insights.

The evolution of computational cardiology reflects a broader transition from handcrafted feature engineering to data-driven representation learning. Early machine learning approaches relied on domain-specific descriptors---such as QRS morphology, interval measurements, and heart rate variability---combined with classical classifiers \cite{abubaker2022detection,kazemi2024advancements}. More recently, deep learning architectures, including Convolutional Neural Networks (CNNs) and Recurrent Neural Networks (RNNs), have demonstrated strong performance in arrhythmia detection, ischemia identification, and risk prediction by learning hierarchical representations directly from raw waveforms \cite{jiang2024automatic, kondo2023prediction}. Despite achieving near-expert performance in controlled benchmarks, these models remain largely task-specific and discriminative, with limited capacity for contextual reasoning, integration of longitudinal clinical data, and robustness across heterogeneous acquisition settings \cite{ding2025advances,wu2025deep}. These challenges motivate the need for more generalizable and clinical knowledge-aware modeling paradigms for cardiovascular intelligence \cite{androulakis2024ai}.

%Computational cardiology has lately transitioned from expert-driven feature engineering to data-driven representation learning, with early ECG analysis pipelines relying on handcrafted descriptors---such as QRS duration, PR interval, QT dispersion, ST-segment deviation, and heart rate variability (HRV)---combined with conventional classifiers including Support Vector Machines (SVMs) and k-Nearest Neighbors (k-NN) \cite{kazemi2024advancements}, which, while effective for well-defined tasks, were limited by their dependence on predefined features and poor adaptability to complex scenarios. The advent of deep learning enabled end-to-end modeling of raw ECG waveforms using convolutional and recurrent neural networks to learn hierarchical temporal-morphological representations directly from data \cite{wu2025deep}, achieving expert-level performance in controlled settings but remaining largely discriminative with limited capacity to incorporate clinical context, longitudinal patient history, or domain knowledge. 

Lately, Large Language Models (LLMs) and Foundation Models (FMs) have emerged as transformative paradigms in artificial intelligence, enabling scalable representation learning and reasoning across various sub-disciplines of health domain \cite{kim2024health, khan2025comprehensive}. Specialized LLMs such as Med-PaLM, Med-Gemma, BioGPT, MEDITRON, and Med42 have demonstrated strong capabilities in clinical question answering, reasoning, report generation, and knowledge synthesis \cite{singhal2023large, singhal2023towards, chen2023meditron, labrak2024biomistral, christophe2024med42}. However, these medical LLMs are predominantly trained on textual corpora—including biomedical literature and electronic health records—and therefore function primarily as knowledge-based reasoning systems rather than direct interpreters of physiological signals. 

Concurrently, a distinct class of \emph{ECG foundation models} has emerged, leveraging large-scale self-supervised learning to model cardiac electrophysiology directly from raw or minimally labeled ECG data \cite{yu2024ecg}. These models capture temporal–morphological patterns and long-range dependencies inherent in ECG signals, enabling robust waveform representation learning and improved generalization across tasks. Recently, researchers have started to explore multimodal and hybrid signal-language architectures that integrate ECG encoders with LLM-based reasoning modules, enabling applications such as automated report generation, multimodal clinical inference, and guideline-aware decision support \cite{chan2024medtsllm, quer2024potential}. Such intertwining of \emph{signal-centric foundation models} and \emph{knowledge-centric medical LLMs} points to a complementary paradigm: ECG foundation models act as waveform analyzers, while medical LLMs provide contextual reasoning based on clinical knowledge.

%At the same time, the locus of ECG intelligence is shifting toward continuous, real-world monitoring enabled by consumer electronics devices such as wearable and edge devices, including smartwatches, sensor patches, holter monitors (for ambulatory monitoring) and mobile health platforms. 
At the same time, ECG intelligence is increasingly moving toward continuous, real-world monitoring facilitated by consumer electronics, particularly wearable and edge devices such as smartwatches, sensor patches, ambulatory Holter monitors, and mobile health platforms.
These systems generate large volumes of physiological data, enabling early detection, personalized risk assessment, and preventive cardiology. However, deploying foundation-scale models in such environments introduces stringent constraints on computation, memory, energy consumption, latency, and data privacy. Large cloud-based models are often impractical for real-time, on-device inference, motivating research into deployment-aware strategies such as quantization, pruning, knowledge distillation, and the development of Small Language Models (SLMs), as well as hybrid edge-cloud architectures \cite{qu2025mobile}. These considerations are particularly critical for ECG applications, where real-time responsiveness and privacy preservation are essential.

Taken together, these developments point toward an emerging paradigm of \emph{agentic cardiovascular intelligence}, in which ECG foundation models and medical LLMs operate in conjunction within integrated pipelines. In such systems, signal encoders extract electrophysiological features from continuous ECG streams, while LLM-based reasoning modules incorporate patient history, symptoms, and clinical guidelines to generate diagnostic insights, prognostic assessments, and actionable recommendations. Despite rapid progress, existing surveys largely examine these components in isolation—focusing either on ECG deep learning, medical LLMs, or edge AI deployment—without providing a unified, deployment-aware perspective. This survey addresses this gap by synthesizing advances across ECG foundation models, medical LLMs, and edge intelligence, and by outlining a translational roadmap toward scalable, trustworthy, and real-time cardiovascular AI systems.

\subsection{Existing Surveys and Research Gap}

%%%%%%%%%%%%%%%%%%%%

%additional survey papers that we may want to add to table 1.
%\cite{Han:2024}, \cite{Lampos:FM}, \cite{Lisicic:2023}, \cite{Lunelli:BenchECG} , \cite{Zhao:Transformer}, \cite{Siontis:2021}, \cite{Ng:2022}, \cite{Qayyum:2025}, \cite{Pantelidis:2023}, \cite{Saini:2022}, \cite{Nechita:2024}, \cite{Maturi:Echocardiography} 

%%%%%%%%%%%%%%%%%%%%

Prior surveys in computational cardiology reflect successive phases of artificial intelligence evolution. Early works \cite{Saini:2022} focused on classical machine learning with handcrafted ECG features, followed by studies \cite{Pantelidis:2023, Qayyum:2025} highlighting convolutional neural network–based deep learning for arrhythmia detection and waveform classification. Clinically oriented reviews \cite{Siontis:2021, Nechita:2024} examined AI applications in specific disease contexts, while more recent efforts have evaluated general-purpose LLMs for cardiology reasoning \cite{novak2023pulse} or proposed taxonomies for biosignal foundation models \cite{Han:2024, Lampos:FM}. Despite these advances, a clear gap remains in systematically understanding how medical LLMs (trained on textual biomedical corpora) can be integrated with ECG foundation models (trained on large-scale waveform data) to enable multimodal cardiovascular intelligence and real-world deployment (see Table \ref{tab:survey_comparison}). Bridging this interdisciplinary gap across natural language processing, physiological signal modeling, and edge AI constitutes the central motivation of this survey.

In contrast to existing surveys that treat these domains in isolation, our work provides a unified and translational perspective. First, we position medical LLMs as knowledge-grounded reasoning backbones that complement ECG foundation models, enabling contextual interpretation, multimodal inference, and clinically actionable decision support rather than direct waveform analysis. Second, we systematically review emerging methodologies spanning self-supervised ECG representation learning, hybrid signal–language architectures, contrastive learning approaches (e.g., MERL, CLOCS, ST-MEM), vision transformer adaptations (e.g., HeartBEiT), and multimodal ECG–language models (e.g., ECG-LM, PULSE, GEM), alongside parameter-efficient adaptation strategies for LLMs \cite{xing2023deeplearningarrhythmia, li2023deeplearningecg}. 

Finally, we provide a deployment-aware roadmap that connects algorithmic advances to practical implementation. This includes analysis of model compression techniques—quantization, pruning, and knowledge distillation—efficient architectures such as small language models and lightweight ECG encoders, and system-level considerations involving edge hardware, real-time inference, and privacy-preserving learning (e.g., federated learning and differential privacy) \cite{zhou2024modelcompression, wang2025edgeai, kumar2025edgedeeplearning}. By further addressing robustness to noisy signals, regulatory constraints, and emerging directions such as agent-driven monitoring and ECG-aware multimodal foundation models, this survey bridges the gap between research innovation and clinical deployment, offering a cohesive roadmap for scalable, trustworthy cardiovascular AI systems.

\begin{table*}[t]
\centering
\caption{Comparison of the Proposed Survey with Existing Survey Papers in Computational Cardiology}
\label{tab:survey_comparison}
\begin{tabular}{@{}lcccccccc@{}}
\toprule
\textbf{Reference} & 
\textbf{\begin{tabular}[c]{@{}c@{}}Classic ML \\ \& Features\end{tabular}} & 
\textbf{\begin{tabular}[c]{@{}c@{}}Deep \\ Learning\end{tabular}} & 
\textbf{\begin{tabular}[c]{@{}c@{}}Foundation \\ Models\end{tabular}} & 
\textbf{\begin{tabular}[c]{@{}c@{}}Medical \\ LLMs\end{tabular}} & 
\textbf{\begin{tabular}[c]{@{}c@{}}Multimodal \\ Integration\end{tabular}} & 
\textbf{\begin{tabular}[c]{@{}c@{}}Agentic \\ AI\end{tabular}} & 
\textbf{\begin{tabular}[c]{@{}c@{}}Edge \\ Intelligence\end{tabular}} \\ \midrule
Saini et al. \cite{Saini:2022} & $\checkmark$ & $\times$     & $\times$     & $\times$     & $\times$     & $\times$     & $\times$     \\
Pantelidis et al. \cite{Pantelidis:2023} & $\checkmark$ & $\checkmark$ & $\times$     & $\times$     & $\times$     & $\times$     & $\times$     \\
Qayyum et al. \cite{Qayyum:2025}     & $\checkmark$ & $\checkmark$ & $\times$     & $\times$     & $\times$     & $\times$     & $\times$     \\
Siontis et al. \cite{Siontis:2021}     & $\checkmark$ & $\checkmark$ & $\times$     & $\times$     & $\times$     & $\times$     & $\times$     \\
Nechita et al. \cite{Nechita:2024}     & $\checkmark$ & $\checkmark$ & $\times$     & $\times$     & $\times$     & $\times$     & $\times$     \\
Novak et al. \cite{novak2023pulse}      & $\times$     & $\times$     & $\times$     & $\checkmark$ & $\times$     & $\times$     & $\times$     \\
Han \& Ding \cite{Han:2024}       & $\times$     & $\checkmark$ & $\checkmark$ & $\checkmark$ & $\checkmark$ & $\times$     & $\times$     \\
Gu et al. \cite{Lampos:FM}         & $\times$     & $\checkmark$ & $\checkmark$ & $\checkmark$ & $\checkmark$ & $\times$     & $\times$     \\ \midrule
\textbf{Our Survey} & $\checkmark$ & $\checkmark$ & $\checkmark$ & $\checkmark$ & $\checkmark$ & $\checkmark$ & $\checkmark$ \\ \bottomrule
\end{tabular}
\end{table*}

\subsection{Survey Methodology}

To provide a rigorous and comprehensive synthesis of this rapidly evolving field, we conducted a structured literature review across major academic databases, including IEEE Xplore, PubMed, Scopus, and Google Scholar. Our search strategy employed Boolean combinations of keywords such as ``Large Language Models,'' ``Foundation Models,'' ``Electrocardiogram,'' ``Multimodal Learning,'' ``Transformer,'' and ``Edge AI.'' We prioritized peer-reviewed journal articles, leading conference proceedings, and high-impact preprints published between 2020 and 2026 to capture recent methodological developments. Selected studies were screened based on their relevance to ECG signal processing, clinical applicability, architectural innovation, and deployment considerations. This process enabled the identification of key research trends and open challenges at the intersection of medical LLMs, cardiovascular intelligence and edge AI.

\subsection{Contributions}

This survey advances the state of the literature by articulating three distinct but interconnected contributions that collectively define a translational perspective on LLM/FM-driven ECG intelligence on the edge:

\begin{itemize}
    \item \textit{Positioning ECG foundation models and medical LLMs for agentic cardiovascular intelligence:} 
    We present a unified system-level perspective in which two distinct classes of foundation models operate synergistically within next-generation cardiology pipelines: (i) ECG foundation models, pre-trained on large-scale unlabeled waveform data, which serve as primary signal interpreters capable of extracting electrophysiological patterns and detecting abnormalities; and (ii) medical LLMs, trained on biomedical literature, electronic health records, and clinical question-answer datasets, which function as knowledge-rich reasoning backbones. 
    We argue that future cardiovascular AI systems will increasingly adopt an agentic paradigm, where these components interact to enable end-to-end intelligence---combining signal-level perception with context-aware reasoning, patient-specific risk assessment, and clinically guided decision support.

    \item \textit{Comprehensive analysis of ECG foundation models and medical LLM-enabled methodologies:} 
    We systematically review the emerging landscape of ECG-specific foundation models alongside LLM-driven approaches, highlighting how self-supervised pretraining on large-scale ECG datasets enables robust waveform representation learning, while LLMs contribute zero-shot and few-shot reasoning capabilities for downstream tasks. 
    The survey integrates these perspectives by examining methodologies that combine both paradigms---spanning arrhythmia detection, automated report generation, prognostics, and multimodal clinical inference---thereby emphasizing the complementary roles of ECG foundation models and medical LLMs in advancing cardiovascular intelligence.

    \item \textit{Deployment-aware roadmap for edge-based ECG AI:} 
    We critically examine model compression techniques---such as quantization, pruning, knowledge distillation, and the development of small language models---as key enablers for translating both ECG foundation models and medical LLMs to wearable and resource-constrained platforms. 
    In doing so, we highlight system-level challenges related to privacy, robustness, real-time performance, and hardware–software co-design, outlining a practical pathway toward scalable, trustworthy, and edge-native cardiovascular AI systems.
\end{itemize}

In short, the survey offers a unified roadmap for designing scalable, explainable, and deployable cardiovascular AI systems that leverage the complementary strengths of ECG foundation models and knowledge-driven medical LLMs.

\subsection{Outline}

The rest of this paper is organized as follows. 
Section II examines medical LLMs as knowledge-based reasoning backbones, highlighting their capabilities in clinical inference and decision support. Section III reviews ECG-specific foundation models, focusing on self-supervised representation learning from large-scale waveform data. Section IV presents a deployment-aware perspective, covering model compression techniques, edge optimization, and challenges related to privacy, robustness, and real-time operation. Section V outlines emerging research directions, including agentic cardiovascular AI systems, while Section VI concludes the paper.

\section{First Enabler for the Agentic ECG AI: Medical LLMs as Reasoning Backbones}

%\subsection{Representation Learning in Computational Cardiology}

%Computational cardiology has lately transitioned from expert-driven feature engineering to data-driven representation learning, with early ECG analysis pipelines relying on handcrafted descriptors---such as QRS duration, PR interval, QT dispersion, ST-segment deviation, and heart rate variability (HRV)---combined with conventional classifiers including Support Vector Machines (SVMs) and k-Nearest Neighbors (k-NN) \cite{kazemi2024advancements}, which, while effective for well-defined tasks, were limited by their dependence on predefined features and poor adaptability to complex scenarios. The advent of deep learning enabled end-to-end modeling of raw ECG waveforms using convolutional and recurrent neural networks to learn hierarchical temporal-morphological representations directly from data \cite{wu2025deep}, achieving expert-level performance in controlled settings but remaining largely discriminative with limited capacity to incorporate clinical context, longitudinal patient history, or domain knowledge. 

%As mentioned earlier, the limitations of ML and DL models has driven a paradigm shift toward integrating high-capacity reasoning systems---particularly medical LLMs---into cardiovascular AI pipelines \cite{androulakis2024ai}, leading to an emerging ecosystem in which ECG-specific foundation models perform signal representation learning while medical LLMs provide higher-level reasoning and contextual understanding. 
This section reviews the research landscape of medical LLMs, some of which are capable of providing the clinical guidelines for the ECG analysis, and thus, make the first set of the enablers for the agentic cardiovascular AI. 

\subsection{Medical LLMs are Knowledge-Centric Reasoning Systems}

Medical LLMs have emerged as powerful tools for biomedical reasoning, clinical documentation, literature synthesis, and medical question answering. However, most of the medical LLMs are fundamentally text-centric and are not designed for direct physiological signal interpretation. Models such as GatorTron, BioGPT, Med-PaLM, Med-PaLM~2, MEDITRON, PMC-LLaMA, BioMistral, Med42-v2, and Clinical Camel \cite{singhal2023large, singhal2023towards, chen2023meditron, labrak2024biomistral, christophe2024med42} are primarily trained on large-scale textual corpora, including electronic health records (EHRs), biomedical publications (PubMed, textbooks), clinical guidelines, and curated clinical question-answer datasets.

Medical LLMs differ in architectural emphasis and downstream capabilities. Encoder-based systems such as GatorTron are optimized for extracting structured clinical information from unstructured narratives, supporting tasks such as concept normalization and relation extraction. Autoregressive models such as BioGPT focus on clinical literature-based text generation and hypothesis formation \cite{luo2022biogpt}, while instruction-tuned systems such as Med-PaLM emphasize multi-step clinical reasoning and standardized medical question answering \cite{singhal2023towards}. Similarly, open-domain biomedical LLMs such as MEDITRON and PMC-LLaMA extend these capabilities to tasks such as diagnosis, and clinical notes summarization.

\subsection{Emerging Multimodal Medical LLMs}

Recent advances in multimodal foundation models aim to bridge the gap between textual reasoning and physiological signal understanding. Architectures such as Med-PaLM~M \cite{tu2024towards} and Med-Gemini introduce cross-modal attention mechanisms that integrate images, text, and structured medical data within a unified framework. Although primarily validated on imaging domains such as radiology and pathology, these models demonstrate preliminary capabilities for ECG-informed reasoning when combined with dedicated signal encoders.
For example, some variants of Med-Gemini can process ECG traces alongside clinical history, achieving moderate performance in arrhythmia and ischemia-related tasks. Importantly, their diagnostic accuracy improves when waveform inputs are supplemented with contextual patient information, highlighting the value of multimodality. Nevertheless, these systems typically underperform compared to specialized ECG foundation models in standalone waveform interpretation, reflecting the inherent mismatch between high-frequency time-series signals and the tokenization paradigms used in text and vision models \cite{ansari2025survey}.
Vision-language biomedical models such as LLaVA-Med have also been explored for ECG image interpretation. However, their performance remains limited without targeted ECG-domain pretraining, reinforcing the need for domain-specific signal representation learning.

\subsection{Hybrid ECG-Language Architectures}

A more effective approach lies in hybrid architectures that explicitly combine ECG foundation models with language model reasoning modules. In this paradigm, dedicated ECG encoders—pretrained on large-scale unlabeled waveform datasets using self-supervised objectives such as masked signal modeling and contrastive learning—extract rich electrophysiological representations. These representations are then aligned with language embeddings to enable downstream reasoning and ECG interpretation tasks.

Recent works such as ECG-LM \cite{yang2025ecglm} and datasets such as ECGInstruct \cite{liu2024pulse} exemplify this approach by linking cardiac signal representations with clinical language understanding. In these architectures, the LLM contributes capabilities beyond classification, including automated report generation, explanation of model predictions, and formulation of differential diagnoses based on clinical guidelines. 

Such hybrid frameworks are particularly relevant for wearable and edge-based monitoring. For instance, an on-device ECG encoder may detect subtle rhythm irregularities in real time, while a lightweight language model synthesizes longitudinal trends, integrates patient-reported symptoms, and recommends appropriate clinical actions. This signal–language collaboration also supports explainability objectives, where attention mechanisms and concept attribution techniques help map predictions to clinically interpretable features \cite{sun2024multimodal}.

\subsection{Role of Medical LLMs in ECG Intelligence}

\begin{figure*}
    \centering
    \includegraphics[width=0.95\linewidth]{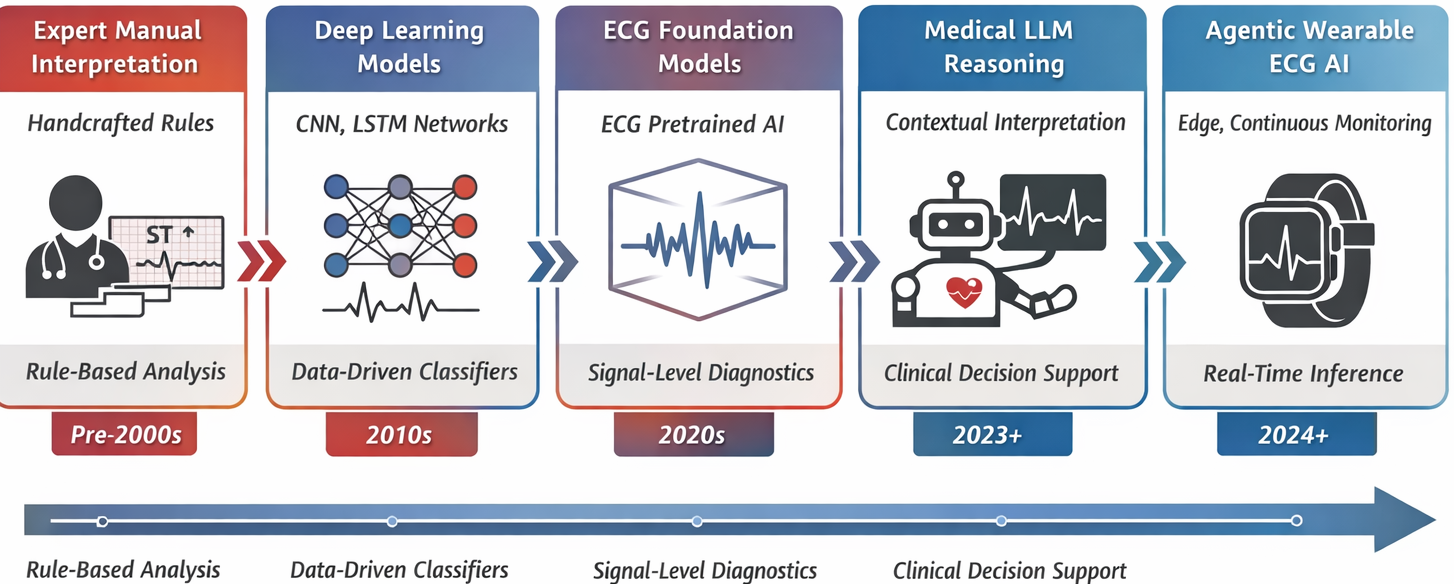}
    \caption{Evolution of AI-based ECG intelligence.}
    \label{fig:ecg_intelligence}
\end{figure*}

From a system-level perspective, medical LLMs should be viewed as reasoning backbones that operate alongside ECG-specific foundation models. Their role can be broadly categorized across a hierarchy of capabilities. At the base level, knowledge-centric LLMs such as GatorTron, BioGPT, and PMC-LLaMA provide access to biomedical knowledge and textual understanding. At the next level, clinical reasoning models such as Med-PaLM, Med42, and MEDITRON enable structured decision-making and multi-step inference. Multimodal generalists, including LLaVA-Med and Med-PaLM~M, extend these capabilities to cross-modal reasoning. Finally, Med-Gemini represents an early instance of ECG-aware multimodal models when combined with signal encoders.
Figure~\ref{fig:ecg_intelligence} illustrates the evolution of ECG intelligence systems, while Table~\ref{tab:medical-llms} summarizes the characteristics of representative medical LLMs and their relevance to ECG reasoning.

Despite their limited ability to process raw ECG waveforms, medical LLMs play a crucial role as knowledge-rich reasoning backbones. They can contextualize outputs from ECG-specific models by incorporating comorbidities, medications, and guideline-based recommendations. For instance, an ECG-derived detection of ST-segment abnormalities can be interpreted in light of patient history and risk factors, enabling more clinically meaningful decision support. Thus, medical LLMs operate at higher cognitive layers of the pipeline, and are complementary to ECG foundation models, which remain essential for robust signal representation learning and real-time inference. Thus, when combined with ECG foundation models, the two systems together form a cohesive agentic framework for next-generation cardiovascular intelligence.
%Importantly, most text-centric models do not directly process ECG waveforms and instead contribute to higher-level functions such as multimodal fusion, contextual reasoning, and report generation. 

%Overall, current evidence indicates that medical LLMs are best understood as knowledge-driven reasoning engines rather than standalone ECG interpreters. Their integration into cardiovascular AI pipelines enables contextual understanding, uncertainty-aware reasoning, and human-interpretable communication. When combined with ECG foundation models, these systems form a cohesive framework for next-generation cardiovascular intelligence.

This paradigm reflects a broader shift from isolated predictive models toward integrated, multimodal intelligence systems. In such pipelines, ECG-specific encoders capture electrophysiological patterns from raw signals, while medical LLMs synthesize clinical knowledge, patient context, and diagnostic reasoning. 
%The following section builds upon this pivot by examining advances in ECG-specific foundation models and LLMs, with emphasis on domain-adaptive pretraining, multimodal fusion, and deployment-aware architectures for scalable and trustworthy cardiology applications.

\begin{comment}

\begin{table}[t]
\caption{Indicative Roles of Medical LLMs Across the ECG Intelligence Pipeline. Here, ``--'' denotes no native capability, while ``Med'', ``High'', and ``Very High'' indicate indicative qualitative relevance within the ECG intelligence pipeline based on architectural design, pretraining modality, and reported downstream use cases.}
\centering
\footnotesize
\setlength{\tabcolsep}{2.5pt}
\begin{tabular}{|l|c|c|c|c|c|}
\hline
\textbf{Model} & \textbf{Signal} & \textbf{Fusion} & \textbf{Reasoning} & \textbf{Reporting} & \textbf{Edge} \\ \hline

GatorTron & -- & Contextual & Med & Med & -- \\ \hline

BioGPT & -- & Literature & Med & High & Med \\ \hline

Med-PaLM & -- & Clinical QA & High & Med & -- \\ \hline

MEDITRON & -- & Evidence & High & Med & Med \\ \hline

BioMistral & -- & Lightweight & Med & Med & High \\ \hline

Med42-v2 & -- & Clinical Query & High & Med & Med \\ \hline

PMC-LLaMA & -- & Knowledge & Med & Med & Med \\ \hline

Clinical Camel & -- & Multilingual & Med & High & Med \\ \hline

Med-PaLM M & ECG Img* & Multimodal & High & Med & -- \\ \hline

LLaVA-Med & ECG Img* & Vision-Lang & Med & Med & Med \\ \hline

Med-Gemini & Partial & Deep Multimodal & Very High & High & -- \\ \hline

\end{tabular}
\label{tab:llm_roles_ecg}
\end{table}

\end{comment}

\begin{table*}[t]
\caption{Representative Medical LLMs and Indicative Relevance to ECG-aware Workflows \cite{yang2022gatortron, luo2022biogpt, singhal2023large, chen2023meditron, tu2024towards, singhal2023towards, labrak2024biomistral, christophe2024med42}}
\centering
\footnotesize
\setlength{\tabcolsep}{3pt}
\begin{tabular}{|l|l|c|c|p{2.4cm}|p{2.8cm}|c|p{3.2cm}|}
\hline
\textbf{Model} & \textbf{Backbone} & \textbf{Params} & \textbf{Context Length} & \textbf{Datasets} & \textbf{Primary Strengths} & \textbf{Access} & \textbf{ECG Suitability (Pretraining-driven)} \\ \hline

GatorTron & BERT & 8.9B & 2k & Large EHR corpora & Clinical NLP, entity/relation extraction & No & Indirect support via clinical note understanding and ECG context integration \\ \hline

BioGPT & GPT-2 & 1.5B & 2k & PubMed abstracts & Biomedical QA, literature synthesis & Yes & Cardiology knowledge grounding; useful for ECG report drafting and explanation \\ \hline

Med-PaLM & PaLM decoder & 540B & 4k & Biomedical corpora + QA & Clinical reasoning, guideline interpretation & No & Can reason over ECG-derived findings for risk assessment or differential diagnosis \\ \hline

MEDITRON & LLaMA-2 & 13B & 2k & Medical literature, QA sets & Literature-grounded reasoning & Yes & Supports ECG interpretation through evidence retrieval and clinical summarization \\ \hline

BioMistral & Mistral 7B & 7B & 2k & Multilingual biomedical text & QA and domain adaptation efficiency & Yes & Lightweight reasoning layer for wearable ECG analytics pipelines \\ \hline

Med42-v2 & LLaMA-3 & 13B & 2k & Clinical query datasets & Decision support and structured responses & Limited & Enables ECG-informed triage suggestions when coupled with signal AI outputs \\ \hline

PMC-LLaMA & LLaMA variant & 13B & 2k & Papers + textbooks & Biomedical QA and synthesis & Yes & Knowledge retrieval for ECG abnormalities and cardiology education \\ \hline

Clinical Camel & Multilingual LLaMA & 7B & 2k & Arabic–English clinical notes & Multilingual clinical NLP & Yes & Useful for ECG report interpretation across linguistic settings \\ \hline

Med-PaLM M & PaLM-E + ViT & 540B+ & 4k & Biomedical text + images & Multimodal reasoning (imaging-centric) & No & Theoretical ECG image understanding after task-specific fine-tuning \\ \hline

LLaVA-Med & Vision-language LLM & 13B & 2k & Medical image–text pairs & Medical VQA (radiology/pathology) & Yes & Can process ECG plots but shows limited accuracy without ECG-domain tuning \\ \hline

Med-Gemini & Gemini multimodal & 70B+ & 4k & Clinical data + signals & Long-context multimodal reasoning & No & Moderate ECG reasoning with adapted encoders and patient history inputs \\ \hline

\end{tabular}
\label{tab:medical-llms}
\end{table*}

%%%%%%%%%%%%%%%%%%%%%%%%%%%%%%%%%%%%%%%%%%%%%%%%%%%%%%%%%%%%%%%%%%%%%%%%%%%%%%%%%%%%%%%

\section{Second Enabler for the Agentic ECG AI: ECG Foundation Models as waveform interpreters}

This section reviews the emerging literature on second main ingredient of agentic ECG AI, i.e., ECG foundation models which are expert physiological waveform interpreters. 

\subsection{Large-Scale Pre-training and ECG Foundation Models}

Large-scale pre-training on unlabeled ECG recordings has emerged as the cornerstone of modern ECG intelligence. ECGFounder \cite{li2025ecgfounder}, trained on 10.77 million ECGs from 1.82 million patients in the Harvard--Emory ECG Database, employs a RegNet-based encoder and positive label augmentation to address incomplete clinical annotations, achieving strong performance across diagnosis, demographic inference, event prediction, and cross-modality tasks on benchmarks including MIMIC-IV-ECG. Similarly, ECGFM \cite{liu2025ecgfm} combines a convolutional encoder with a transformer decoder and task-specific heads, jointly optimizing contrastive predictive learning, normal–abnormal classification, and diagnostic text generation to enable interpretable clinical deployment.

Self-supervised foundation modeling strategies have also drawn inspiration from speech processing. HuBERT-ECG \cite{hubert2024ecg}, pre-trained on 9.1 million ECGs across 164 cardiovascular conditions, employs masked cluster prediction to iteratively refine representations, achieving strong transferability across diagnostic and prognostic tasks. Open-weight efforts such as ECG-FM \cite{mckeen2024ecgfm} further promote reproducibility, integrating wav2vec~2.0-style masking, contrastive multi-segment coding (CMSC), and random lead masking within a hybrid convolutional–transformer architecture that demonstrates high label efficiency for tasks such as left ventricular ejection fraction prediction.

\subsection{Self-Supervised Learning Approaches}

Contrastive and masked modeling approaches have enabled ECG representation learning without extensive annotations. \cite{tapclr2024} presents temporally-augmented patient-contrastive learning (TA-PCLR) method that leverages patient identity, temporal proximity, and signal augmentation to improve downstream classification on PTB-XL, while large multi-continent pre-training cohorts enhance external validity. In \cite{kiyasseh2021clocs}, Kiyasseh et. al. propose contrastive learning of cardiac signals (CLOCS) method that exploits temporal adjacency and multi-lead spatial invariance through contrastive multi-lead coding (CMLC) and hybrid CMSMLC, consistently outperforming baseline self-supervised learning methods such as simple contrastive learning (SimCLR) and bootstrap your own latent (BYOL) on datasets such as Chapman and PhysioNet 2020.
Complementing contrastive methods, \cite{na2024stmem} presents spatio-temporal masked electrocardiogram modeling (ST-MEM) method that reconstructs masked multi-lead signals using lead-specific attention to capture inter-lead spatial dependencies and intra-lead temporal dynamics. Its adaptability to incomplete lead configurations highlights practical clinical relevance where full 12-lead recordings may not be available.

\subsection{Multimodal ECG Foundation Models}

\textit{Cross-Modal Alignment:}
To mitigate annotation scarcity, cross-modal alignment strategies map ECG representations into semantic language spaces. Retrieval-augmented zero-shot frameworks \cite{yu2023zero} combine fiducial feature extraction with structured prompting to enable general-purpose LLMs such as GPT-3.5 and LLaMA-2 to diagnose arrhythmias and sleep apnea with competitive macro-precision. \cite{liu2024merl} presents multimodal ECG representation learning (MERL) method which aligns ECG waveforms and reports through contrastive Cross-Modal Alignment (CMA) and latent-space Uni-Modal Alignment (UMA), enabling strong zero-shot classification across six benchmarks including PTB-XL and CPSC2018. They further improve performance and reduce hallucinations by enhancing the clinical knowledge of their model through prompt engineering.
Another work \cite{liu2024etp} presents ECG-text projection (ETP) method, which aligns ResNet18 ECG encodings with BioClinicalBERT representations for robust zero-shot inference. \cite{tang2024electrocardiogrammeta} proposes a meta-learning framework that combines self-supervised ECG encoders with frozen decoder-only LLMs such as Gemma. Finally, in \cite{qiu2023transfer}, Qiu et. al. utilize transfer learning using optimal transport alignment to enable hierarchical mapping between transformer-derived ECG features and language embeddings. 

\textit{Clinical Knowledge Enhancement of LLMs and FMs:}
There have been quite some efforts to reduce hallucinations in LLMs and foundation models by means of clinical knowledge enhancement of such models via prompting and instruction tuning. In \cite{wang2024ked}, Wang et. al. present the knowledge-enhanced ECG Diagnosis foundation model (KED), which is trained on 800,000 ECGs, and integrates signal–text–label contrastive objectives to achieve expert-level zero-shot diagnosis across diverse populations and unseen conditions.
Building upon MERL, \cite{kmerl2025} presents knowledge-enhanced MERL which introduces lead-specific tokenization and structured knowledge extraction from reports, improving performance in zero-shot and linear probing tasks, particularly with partial-lead inputs. Another work \cite{melp2025} presents multi-scale ECG language pretraining (MELP) method that further refines alignment by operating at token, segment, and rhythm levels. In \cite{fgclep2025}, authors propose fine-grained contrastive learning for ECG pretraining (FG-CLEP) method that generates and validates missing waveform features using LLM guidance.
Finally, \cite{supreme2024} presents supervised pre-training for multimodal ECG (SuPreME) method that offers an alternative by fusing ECG signals with structured cardiac queries mapped to terminologies such as systematized nomenclature of medicine clinical terms (SNOMED CT), unified medical language system (UMLS), and standard communications protocol for computer-assisted electrocardiography (SCP-ECG). Its cardiac fusion network achieves competitive performance without extensive augmentation while improving interpretability.

\textit{Vision Transformer Architectures:}
Vision transformer adaptations have demonstrated strong capability for global dependency modeling. HeartBEiT \cite{vaid2023heartbeit}, pre-trained on 8.5 million ECGs using masked image modeling, treats ECGs as patch-tokenized images processed via a DALL-E tokenizer. With approximately 86 million parameters, it surpasses convolutional neural networks such as ResNet-152 and EfficientNet-B4 in low-data ST-elevation myocardial infarction detection while providing more clinically meaningful saliency localization. 

\textit{ECG-Language Models:}
Multimodal ECG-language models increasingly support clinical reporting and question answering. 
Earlier generative pipelines such as SignalGPT \cite{liu2023biosignal} convert signal features into textual descriptors refined through reinforcement learning, while ECG semantic integration frameworks \cite{yu2024ecg} combine ConvNeXt encoders with BioLinkBERT to power retrieval-augmented cardiology assistants. Conversational diagnostic systems including ECG-Chat \cite{zhao2024ecg} employ Vision Transformers and Graph Retrieval-Augmented Generation (GraphRAG). The work \cite{wan2024electrocardiogram} presents multimodal electrocardiogram instruction tuning (MEIT) method, which utilizes low-rank adaptation-based LLaMA and Mistral backbones to improve report generation metrics on MIMIC-IV-ECG.
ECG-LM \cite{yang2025ecglm} aligns signal embeddings with textual representations from BioMedGPT-LM-7B, using synthetic text–ECG pairs derived from medical guidelines to address data scarcity. Q-HEART \cite{qheart2025} incorporates instruction-tuned reasoning with retrieval of similar ECG cases to improve contextual understanding. The work \cite{gem2025} presents grounded ECG understanding model (GEM), which integrates time-series, image, and text modalities to provide feature-based diagnostic explanations.

\textit{Visual ECG Interpretation:}
Instruction-tuned multimodal datasets have enabled interpretation of printed ECGs. ECGInstruct \cite{liu2024pulse}, a dataset containing over one million synthesized ECG images with realistic artifacts, supports diverse reasoning tasks. The PULSE model fine-tuned on this dataset achieves state-of-the-art performance on ECGBench dataset, outperforming general multimodal large language models. 
%by 15--30\% accuracy and demonstrating practical relevance in settings lacking access to raw signals.

\subsection{Prognostics, Risk Stratification, and Biomarker Estimation}

Foundation models extend ECG utility beyond diagnosis toward prognostic inference. Applications include prediction of reduced ejection fraction, risk of heart failure, mortality estimation, and outcome forecasting. MedTsLLM \cite{chan2024medtsllm} maps physiological time-series patches into language embedding spaces for anomaly detection and segmentation, while dual-attention frameworks initialized via LLM-guided pre-training \cite{chen2024large} align ECG patterns with ClinicalBERT semantics to identify temporal markers of cardiac risk.

Emerging biomarker estimation studies demonstrate non-invasive physiological inference. Instruction-tuned LLaMA-3 models estimate blood pressure from wearable ECG and photoplethysmography (PPG) signals \cite{liu2024large}, whereas zero-shot analyses using multimodal LLMs such as Idefics9B explore dehydration detection and chronological age estimation from single-lead ECGs \cite{perzhilla2025situ, ali2025peft}.

\begin{figure*}
    \centering
    \includegraphics[width=1\linewidth]{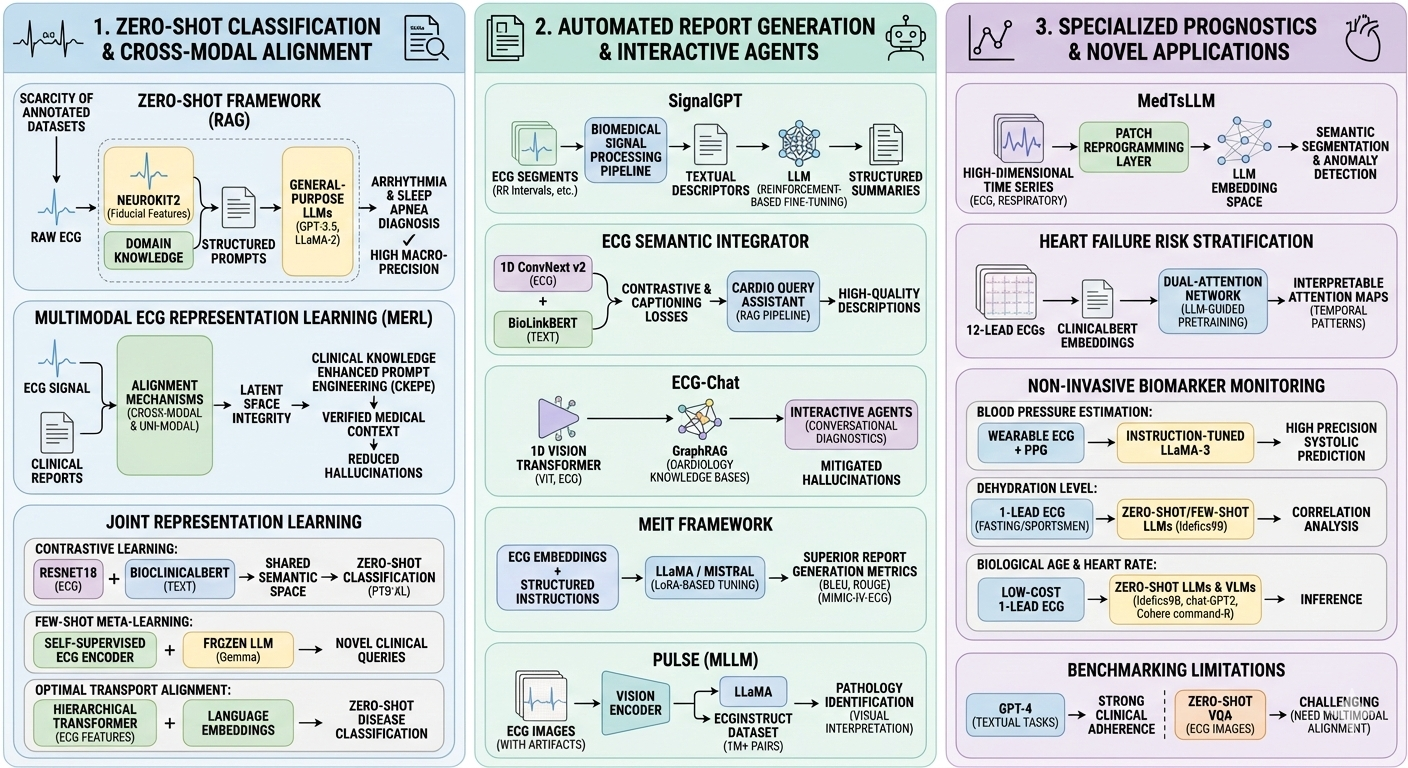}
    \caption{The current research landscape of the use of Foundation models and LLMs for ECG analysis.}
    \label{fig:llm-ecg2}
\end{figure*}

Figure~\ref{fig:llm-ecg2} and Table \ref{tab:ecg_models_summary} summarize the research landscape of ECG foundation models in visual and tabular formats, respectively.

\subsection{Discussion}

%Across foundation models, architectural diversity persists, spanning convolutional encoders in ECGFM and ECG-FM, transformer-based temporal modeling in HuBERT-ECG, and hybrid designs that balance efficiency and expressivity.

\textit{Clinical Deployment Considerations}:
Real-world deployment of ECG foundation models necessitates robustness to imperfect recordings. TolerantECG \cite{tolerantecg2024} is an ECG foundation model that utilizes self-distillation with no labels (DINO) method to improve resilience to noise and missing leads. Multimodal contrastive extensions such as MERL-ECHO \cite{merlecho2024} align ECGs with echocardiography reports for zero-shot structural heart disease prediction, achieving moderate but clinically meaningful performance and demonstrating the value of cross-modality learning.

\textit{Performance Benchmarking}:
Though ECG foundation model KED \cite{wang2024ked} achieves cardiologist-level zero-shot performance, MERL \cite{liu2024merl} with prompt enhancement surpasses supervised baselines trained on limited data, and PULSE \cite{liu2024pulse} significantly improves multimodal reasoning accuracy. Nevertheless, evaluations \cite{novak2023pulse, seki2024assessing} highlight persistent challenges, including limited visual question answering accuracy on ECG images and limited multimodal alignment.

\textit{Pre-training Objectives}:
Pre-training strategies have evolved from single-objective contrastive learning toward hybrid frameworks combining masked reconstruction, classification, and generative objectives. Evidence across studies suggests that integrating complementary objectives enhances representation richness and downstream adaptability. Cross-modal alignment further improves signal representations in clinical semantics, enabling flexible generalization.

\textit{Multimodal Integration and Semantic Capabilities}:
Integration of ECG signals with textual and structured medical knowledge unlocks capabilities such as zero-shot classification via prompt-defined labels, automated narrative report generation, and natural language explanations supporting interpretability. These developments reflect a broader shift toward knowledge-aware diagnostic systems that emulate expert reasoning.

\textit{Public ECG Datasets and Implications for ECG Foundation Models Research}:
Progress in ECG foundation modeling is closely tied to dataset availability. Large repositories such as HEEDB, multi-country cohorts used for HuBERT-ECG, and MIMIC-IV-ECG support large-scale pre-training, while benchmarks including PTB-XL and CPSC 2018 remain essential for evaluation. Emerging resources such as ECG-QA introduce conversational reasoning formats. However, efforts are needed in synthetic data generation, and longitudinal, multimodal data acquisition via wearables across population groups. Table~\ref{tab:ecg_datasets} summarizes currently existing representative ECG datasets.

\textit{Clinical Applications and Impact}:
ECG foundation models have demonstrated effectiveness across arrhythmia detection, myocardial infarction diagnosis, structural abnormality identification, conduction disorder classification, and ischemia recognition. Their multi-label inference capability enables holistic cardiac assessment from a single recording. Integration into clinical workflows can accelerate triage, reduce reporting burden, support training, and provide decision support while maintaining clinician oversight.
Some additional important considerations for ECG foundation models include privacy-preserving large-scale training, improved explainability through saliency visualization and uncertainty quantification, and robustness to demographic shifts, device variability, and signal artifacts. 

\begin{table*}[t]
\centering
\caption{Summary of State-of-the-Art ECG Foundation Models and LLM-Driven Architectures}
\label{tab:ecg_models_summary}
\renewcommand{\arraystretch}{1.3}
\begin{tabularx}{\textwidth}{l p{2.2cm} p{4.5cm} p{3.5cm} X}
\hline
\textbf{Category} & \textbf{Model} & \textbf{Core Methodology / Innovation} & \textbf{Primary Dataset / Scale} & \textbf{Key Clinical Contribution} \\ \hline

\textbf{Large-Scale} & \textbf{ECGFounder} \cite{li2025ecgfounder} & RegNet with positive label augmentation for missing annotations. & 10.77M ECGs (HEEDB) & State-of-the-art general-purpose ECG encoder for 150+ categories. \\
\textbf{Foundation} & \textbf{HuBERT-ECG} \cite{hubert2024ecg} & Iterative clustering and masked modeling adapted from speech. & 9.1M ECGs (Multi-country) & Robust detection of 164 conditions with high transferability. \\
\textbf{Models} & \textbf{ECG-FM} \cite{mckeen2024ecgfm} & WCR: wav2vec 2.0 + CMSC + Random Lead Masking. & 1.5M ECGs (MIMIC-IV) & Open-weight model; exceptional performance in low-data regimes. \\ \hline

\textbf{Spatio-Temporal} & \textbf{ST-MEM} \cite{na2024stmem} & Lead-specific masked modeling for spatio-temporal reconstruction. & - & Explicitly captures inter-lead and intra-lead relationships. \\
\textbf{Learning (SSL)} & \textbf{CLOCS} \cite{kiyasseh2021clocs} & Patient-specific spatial and temporal contrastive learning. & Chapman, PhysioNet 2020 & Eliminates semantic distortion from traditional data augmentation. \\ \hline

\textbf{Cross-Modal} & \textbf{MERL} \cite{liu2024merl} & Cross-modal alignment with clinical RAG and knowledge prompts. & PTB-XL, CPSC 2018 & Robust zero-shot classification; mitigates LLM hallucinations. \\
\textbf{Alignment} & \textbf{K-MERL} \cite{kmerl2025} & Spatial tokenization + structured entity extraction from reports. & - & Superior zero-shot performance with partial/missing leads. \\
& \textbf{ETP} \cite{liu2024etp} & Contrastive alignment of ResNet-ECG with BioClinicalBERT. & PTB-XL & Bridges unannotated signals with shared semantic space. \\ \hline

\textbf{Generative \&} & \textbf{ECG-LM} \cite{yang2025ecglm} & Aligns dedicated ECG encoder with BioMedGPT-7B. & Clinical conversations & First multimodal LLM for ECG-related Q\&A and reasoning. \\
\textbf{Interactive} & \textbf{GEM} \cite{gem2025} & Feature-grounded conversational cardiology assistant. & Multimodal benchmark & Explicitly ties diagnoses to granular waveform evidence. \\
\textbf{Agents} & \textbf{PULSE} \cite{liu2024pulse} & Instruction-tuned MLLM via the ECGInstruct dataset. & 1M+ image-text pairs & Interprets printed/digital ECG images with real-world artifacts. \\ \hline

\textbf{Specialized} & \textbf{MedTsLLM} \cite{chan2024medtsllm} & Patch reprogramming layer for time-series into LLM space. & ECG + Respiratory & Robust semantic segmentation and anomaly detection. \\
\textbf{Prognostics} & \textbf{MERL-ECHO} \cite{merlecho2024} & Contrastive alignment of 12-lead ECG with Echo reports. & 45,016 ECG-Echo pairs & Zero-shot prediction of structural diseases (e.g., LVEF). \\
& \textbf{TolerantECG} \cite{tolerantecg2024} & DINO-based self-distillation with feature retrieval. & - & Ensures diagnostic robustness against noise and missing leads. \\ \hline
\end{tabularx}
\end{table*}

%Fig. \ref{fig:llm-ecg2} provides a pictorial summary of this section. 

\begin{table*}[t]
\centering
\scriptsize
\renewcommand{\arraystretch}{1.1}
\setlength{\tabcolsep}{2.8pt}
\caption{Publicly available ECG datasets, with characteristics relevant for FM/LLM training and evaluation.}
\label{tab:ecg_datasets}
\resizebox{\textwidth}{!}{
\begin{tabular}{|p{2.1cm}|p{1.0cm}|p{1.0cm}|p{2.8cm}|p{3.0cm}|p{1.9cm}|p{1.8cm}|p{2.1cm}|}
\hline
\textbf{Dataset} & \textbf{Size} & \textbf{Subjects} & \textbf{Recordings / Duration} & \textbf{Acquisition Context / Device} & \textbf{Annotations} & \textbf{Additional Signals} & \textbf{Typical Downstream Tasks} \\
\hline
MIT-BIH Arrhythmia Database \cite{932724} & 48 records & 47 & 2 × 30-min ECG/channel & Ambulatory clinical ECG, 360 Hz & Beat-level expert annotations & None & Arrhythmia detection, beat classification, signal segmentation \\
\hline
MIT-BIH Atrial Fibrillation DB \cite{moody1983new} & 25 long records & 25 & 10+ hr ECG & Ambulatory ECG, 250 Hz & Rhythm labels for AF episodes & None & AF detection, rhythm classification \\
\hline
MIT-BIH Supraventricular Arrhythmia DB \cite{mark1990bih, greenwald1990improved} & 78 records & ~78 & 30-min ECG & Ambulatory ECG & Beat annotations for SV arrhythmias & None & SV arrhythmia detection and classification \\
\hline
Long-Term ST Database \cite{jager2003long} & 86 recordings & 80 & Hours-long ECG & Ambulatory stress/ischemic tests & ST segment events & None & Ischemia, ST change detection \\
\hline
ECG-ID Database \cite{lugovaya2005biometric} & ~90 subjects & 90 & 2–20 short ECGs per subject & Short single-lead ECG & Annotated identity labels & None & Biometric identification, pattern representation  \\
\hline
PTB Diagnostic ECG \cite{bousseljot1995nutzung} & 549 ECGs & 294 & Multi-lead clinical ECG & Clinical ECG (15 leads) & Cardiologist summaries & None & MI/heart disease diagnosis, multi-lead analysis \\
\hline
PTB-XL \cite{PhysioNet-ptb-xl-1.0.1, wagner2020ptb} & 21801 ECGs & 18869 & 10 s 12-lead ECGs & Clinical ECG (500–1000 Hz) & Up to two cardiologist labels + metadata & Demographics, rhythms & Arrhythmia classification, phenotyping, multi-label tasks \\
\hline
CPSC 2018 \cite{liu2018open} / CPSC-Extra & ~10000+ ECGs & ~10000+ & 12-lead ECGs & Clinical ECG (challenge) & Lead-wise arrhythmia labels & None & Arrhythmia detection, diagnostics \\
\hline
INCART 12-lead Database \cite{tihonenkost} & 75 records & ~32 & 30-min 12-lead ECG & Clinical ECG & Beat/annotation files & None & Multi-lead arrhythmia \\
\hline
European ST-T Database \cite{taddei1992european} & 90 ECGs & ~90 & Up to 2 hrs & Clinical ECG & Expert ST change labels & None & ST analysis, ischemia detection  \\
\hline
AHA ECG Database (sample) \cite{PhysioNet-ecg-arrhythmia-1.0.0, zheng2020optimal} & Small subset & N/A & Limited ECG segments & Clinical ECG & Rhythm-level annotations & None & Pilot diagnostic algorithm validation  \\
\hline
Fetal ECG Database \cite{andreotti2016open} & 55 ECGs & 1 mother & Maternal + fetal ECG & Clinical/NI ECG acquisition & None / partial labels & None & Fetal ECG extraction and characterization  \\
\hline
LUDB (Lobachevsky ECG) \cite{PhysioNet-ludb-1.0.1, kosonogov2020ludb} & 200 ECGs & 200 & 10 s 12-lead ECGs & Clinical ECG & Waveform delineations & None & Delineation, segmentation, lead-wise tasks  \\
\hline
ECG-QA (ECG + QA) \cite{oh2023ecg} & 70 Q/A templates & ECG signal subset & ECG + expert questions & ECG with interpretation context & Expert-validated Q/A & None & Clinical ECG QA, interpretability  \\
\hline
Icentia11k Continuous ECG \cite{tan2019icentia11k, PhysioNet-icentia11k-continuous-ecg-1.0} & ~11000 subjects & ~11000 & Continuous raw ECG & Wearable ECG & Beat annotations & None & Representation learning, unsupervised ECG modeling  \\
\hline
Sleep ECG (MIT-BIH PSG) \cite{ichimaru1999development} & 18 subjects & 18 & Sleep ECG + EEG & Clinical polysomnography & Sleep/arrhythmia labels & EEG & Sleep/cardiac interplay modeling  \\
\hline
VitalDB (cardiac signals) \cite{lee2022vitaldb, PhysioNet-vitaldb-1.0.0} & ~6388 patients & ~6388 & ECG + multi-signals & ICU (clinical context) & ICU tags & ECG + vitals & Multimodal time-series modeling  \\
\hline
\end{tabular}
}
\end{table*}

%%%%%%%%%%%%%%%%%%%%%%%%%%%%%%%%%%%%%%%%%%%%%%%%%%%%%%%%%%%%%%%%%%%%%%%%%%%%%

\section{Realizing Agentic ECG AI on the Edge: Methodological Enablers and Challenges}

The deployment of AI models such as LLMs, FMs, SLMs, DL and ML models for ECG analysis on consumer/edge devices represents a critical step toward achieving real-time, privacy-preserving, and energy-efficient cardiovascular intelligence. While large-scale models offer strong reasoning and multimodal capabilities, their computational and memory demands hinder direct deployment on wearable and embedded platforms. Consequently, a combination of algorithmic innovations, model compression techniques, efficient architectures, edge-cloud hybrid systems, and hardware–software co-design strategies has emerged to enable practical edge intelligence.

\subsection{Model Optimization and Edge Deployment Ecosystem}

\subsubsection{Quantization Techniques}

Quantization is a fundamental approach for reducing model size and computational complexity by representing parameters and activations in low-precision formats. Recent work demonstrates that aggressive quantization can retain high accuracy when combined with appropriate optimization strategies. LoftQ \cite{li2023loftq} integrates quantization with low-rank adaptation (LoRA) to align fine-tuning updates with quantized weights, enabling robust 2-bit and 4-bit adaptation. SmoothQuant \cite{xiao2023smoothquant} redistributes activation scaling difficulty to weights, enabling efficient INT8 inference for Transformer-based architectures such as LLaMA with approximately twofold memory reduction. OmniQuant \cite{shao2023omniquant} extends this approach using learnable clipping and equivalent transformations for improved robustness across architectures.

At ultra-low precision, QuIP \cite{chee2024quip} employs orthogonal transformations to enable accurate 2-bit quantization, while methods such as Additive Quantization of Language Models (AQLM) \cite{egiazarian2024extreme} jointly optimize transformer blocks for improved compression–accuracy trade-offs. Parameter-efficient approaches such as Quantization-Aware Low-Rank Adaptation (QA-LoRA) method \cite{xu2023qa} further enable low-precision adaptation, making them suitable for personalized ECG monitoring systems on edge devices.

%From an application perspective, quantization has been widely explored for ECG models. Holda et al. \cite{holda2025ann} showed that quantization reduces model size but does not always translate into proportional runtime improvements due to hardware dependencies. In contrast, hybrid pipelines combining quantization with other techniques often yield better results. Post-training quantization (PTQ) is commonly used for rapid deployment, whereas quantization-aware training (QAT) improves accuracy retention for safety-critical ECG applications \cite{xu2026cardiolike}.

From an application perspective, there are a handful of works that do quantization for compressing ECG classification models for edge deployment. However, its impact on inference latency is not always proportional to the reduction in model size. Holda et al. \cite{holda2025ann} demonstrate that while quantization significantly shrinks model footprint, the resulting runtime improvements depend heavily on hardware characteristics such as memory bandwidth and instruction set support, sometimes yielding only marginal speedups. Consequently, hybrid pipelines that combine quantization with complementary compression techniques often achieve superior efficiency gains. Regarding quantization strategy, post-training quantization (PTQ) enables rapid deployment with minimal calibration data, making it suitable for many edge scenarios \cite{zhu2025real}. For safety-critical ECG applications where diagnostic accuracy is paramount, quantization-aware training (QAT) is preferred as it better preserves classification fidelity by simulating quantization errors during model training \cite{zhang2024energy,xu2026cardiolike}.

\subsubsection{Pruning, Sparsity, and Hybrid Compression}

Pruning complements quantization by removing redundant parameters, thereby reducing both model size and computational cost. SparseGPT \cite{frantar2023sparsegpt} formulates pruning as a sparse regression problem, achieving up to 60\% sparsity in large models without retraining, while SparseLLM \cite{bai2024sparsellm} extends this approach to higher sparsity regimes. Structured pruning methods such as Structured Optimal Brain Pruning (SoBP) \cite{wei2024structured} identify optimal pruning patterns with minimal fine-tuning, and KVPruner \cite{lv2024kvpruner} targets attention key-value (KV) cache redundancy, reducing memory usage and improving throughput.

Hybrid approaches integrating pruning, quantization, and knowledge distillation have proven particularly effective. \cite{zafrir2021prune} proposes Prune Once for All (PruneOFA) method that combines pruning with distillation to preserve performance during transfer learning. In ECG-specific applications, Chang et al. \cite{chang2022ecg} achieved a 91-fold compression for atrial fibrillation detection with minimal accuracy loss, while Lee et al. \cite{lee2022compression} demonstrated compression factors up to 10,000× for arrhythmia classification using pruning, quantization, and clustering. Hybrid pipelines \cite{compression2025hybrid} where sequential pruning is followed by quantization provide optimal accuracy-efficiency trade-offs.

Knowledge distillation plays a crucial role in these pipelines by transferring knowledge from large teacher models to smaller student models. Studies such as \cite{holda2025ann} highlight that distillation consistently yields the smallest accuracy degradation, making it particularly suitable for edge-deployed ECG intelligence systems.

\subsubsection{Small Language Models and Efficient Architectures}

An alternative to compressing large models is the design of small language models (SLMs) optimized for efficiency. Models such as Phi-3-mini and TinyLlama \cite{javaheripi2023phi, zhang2024tinyllama} demonstrate that carefully curated datasets and hardware-aware training can deliver strong reasoning performance at significantly reduced parameter counts. In healthcare, SLMs provide a favorable balance between performance and efficiency, particularly for real-time ECG monitoring on edge devices.

HealthSLM-Bench \cite{healthslm2025} systematically evaluates SLMs such as Phi-3-mini, TinyLlama, and Gemma2 across healthcare monitoring tasks, demonstrating that these models can operate within smartphone constraints (e.g., iPhone-class devices) while maintaining competitive performance. Surveys such as \cite{subramanian2025slm} further emphasize that SLMs reduce energy consumption by 10–100× compared to large models and enable subsecond inference, making them suitable for time-critical medical applications.

Recent studies \cite{bjorkdahl2024towards, wang2024efficient} demonstrate that multimodal SLMs can be adapted to jointly process ECG signals and clinical text, enabling on-device prediction of cardiac events. Architecturally, emerging alternatives such as state-space models (e.g., Mamba) \cite{hu2025adaptive} and gated linear attention Transformers provide linear or near-linear scaling with sequence length, making them well suited for long-duration ECG recordings. 
%Streaming inference techniques and efficient attention mechanisms further enhance robustness and reduce latency \cite{fan2024towards}.

\subsubsection{Tiny ML for Microcontroller-Based ECG Monitoring}

TinyML framework enables deployment of ML models on ultra-low-power microcontrollers with strict memory and energy constraints. This paradigm is particularly relevant for wearable ECG monitoring, where continuous operation and privacy are essential.

Kim et al. \cite{kim2024tinyml} developed a TinyML-based ECG classification system achieving 97.34\% accuracy with a quantized model of only 143 kilobytes, demonstrating real-time monitoring capability. Busia et al. \cite{busia2024tiny} proposed a Tiny Transformer model with only 6,649 parameters, achieving 98.97\% accuracy and extremely low energy consumption (0.09 millijoules per inference). Similarly, Alimbayeva et al. \cite{alimbayeva2024wearable} demonstrated a wearable ECG system combining low-power hardware with optimized convolutional neural networks. Complementing these approaches, Cioflan et al. \cite{cioflan2025nanohydra} introduced NanoHydra that employs lightweight binary random convolutional kernels for feature extraction. Evaluated on the ECG5000 dataset using the ultra-low-power GAP9 microcontroller, NanoHydra achieves 94.47\% accuracy, while requiring only 0.33 ms to classify a one-second ECG signal. With an energy consumption of 7.69 µJ per inference, it is 18× more efficient than prior works, enabling a device lifetime exceeding four years in smart wearable applications.

The concept of micro-trainers ($\mu$-trainers) introduced by Huang et al. \cite{huang2025microtrainers} enables on-device personalization through fine-tuning and meta-learning. Comprehensive surveys \cite{tinyml2025survey, abadade2023comprehensive} highlight that TinyML requires holistic co-design of model architecture, compression strategy, and hardware platform to achieve optimal performance.

\subsubsection{Federated Learning for Privacy-Preserving ECG Intelligence}

Federated learning enables collaborative model training across distributed edge devices without sharing raw patient data, addressing key privacy concerns in healthcare. Alkahtani et al. \cite{alkahtani2025federated} proposed a secure edge-AI framework integrating federated learning, blockchain, and homomorphic encryption, achieving high accuracy with minimal latency overhead. Rajagopal et al. \cite{rajagopal2023fedsdm} introduced FedSDM for smart ECG decision-making in edge–fog–cloud environments, while Gao et al. \cite{gao2025federated} demonstrated federated fine-tuning of LLMs using heterogeneous quantization and LoRA.

Additional studies \cite{mohammed2023energy, fedhealth2024, fedprivacy2025} highlight the importance of energy efficiency, data heterogeneity, and security challenges in federated healthcare systems. Although some works focus on related applications such as seizure detection \cite{aminifar2024privacy}, they reinforce the importance of combining federated learning with compression techniques (e.g., quantization and sparsification) to reduce communication overhead.

\subsubsection{Software Frameworks and Hardware Co-Design}

Efficient edge deployment requires tight integration between software frameworks and hardware accelerators. Lightweight runtimes such as TensorFlow Lite for Microcontrollers and ExecuTorch enable inference on microcontroller units using optimized kernels (e.g., CMSIS-NN), while TensorFlow Lite, PyTorch Mobile, and ONNX Runtime support more capable edge devices. Compiler frameworks such as Apache TVM and MLIR/IREE perform graph-level optimizations including operator fusion and hardware-aware scheduling.

Hardware platforms range from ultra-low-power microcontrollers (e.g., ESP32-S3, Ambiq Apollo) to advanced system-on-chip (SoC) devices with neural processing units, such as Qualcomm Snapdragon Wear platforms and Apple Neural Engine-enabled processors. Edge computing frameworks leveraging 5G connectivity \cite{spicher2022edge, batool2025realtime} further enable real-time ECG analysis with reduced latency.

Application-specific systems demonstrate practical feasibility. Honda et al. \cite{honda2024multimodal} developed a multimodal sensor patch integrating ECG and other vital signals with smartphone-based edge processing. Suganya et al. \cite{suganya2025deeplearning} proposed an interoperable edge framework combining federated learning and knowledge distillation, while the role of quantization and distillation in deploying Transformer and state-space models on microcontrollers is also important.

\begin{figure*}
    \centering
    \includegraphics[width=1\linewidth]{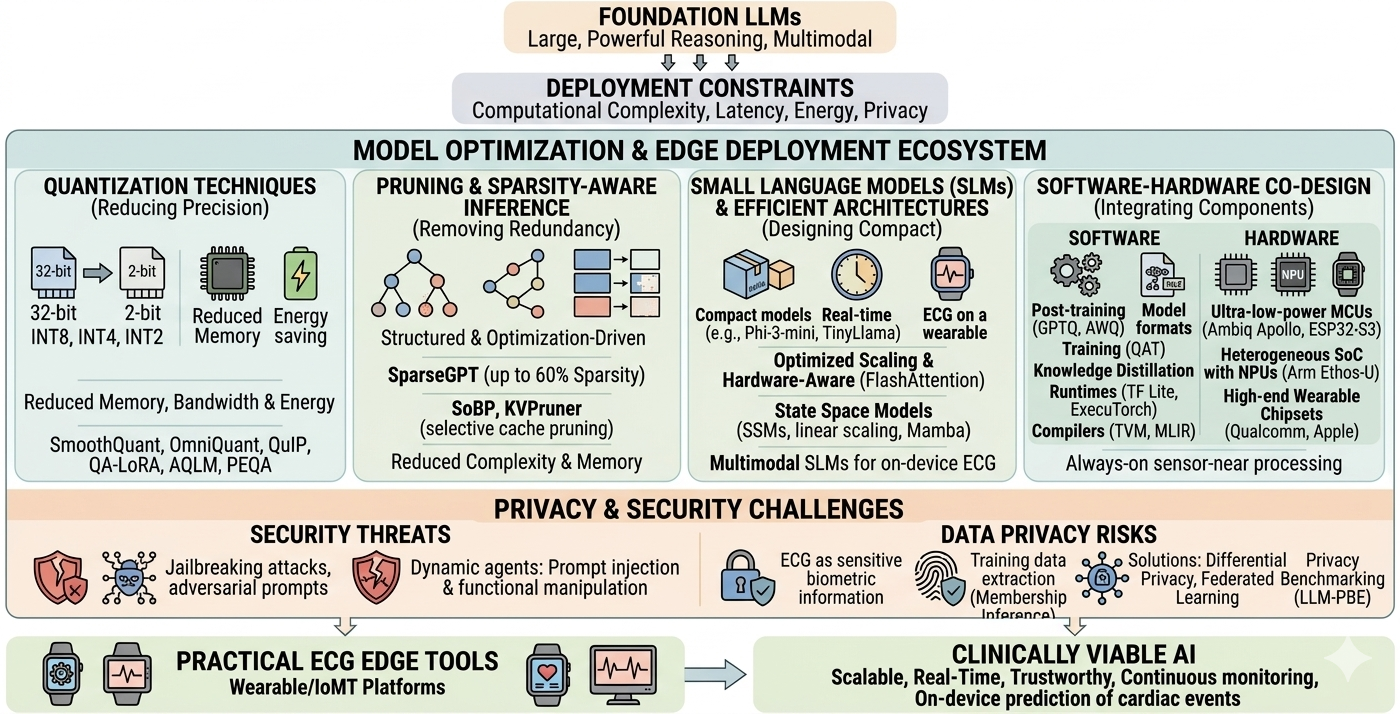}
    \caption{Model optimization, hardware-software co-design, and privacy-aware AI techniques to realize Agentic ECG intelligence on the edge.}
    \label{fig:edge_AI_ECG}
\end{figure*}

Figure~\ref{fig:edge_AI_ECG} provides a compact graphical overview of the key methodologies that have the potential to enable AI-based ECG intelligence at the edge.

\subsubsection{Clinical Deployment and Real-World Applications}

As of now, a handful of works have demonstrated the viability of edge-based ECG intelligence. CardioHelp \cite{utsha2024cardiohelp} enables real-time ECG analysis directly on smartphones without cloud connectivity. Other edge-based healthcare applications, including fall detection and multimodal monitoring systems \cite{paramasivam2024fall, batool2025realtime}, illustrate the broader applicability of edge artificial intelligence techniques, although their discussion here is primarily to contextualize system-level design considerations.
Systematic reviews \cite{edgeml2024systematic, jcloud2024integration} report that edge-enabled healthcare systems achieve high accuracy while significantly reducing data transmission requirements, through local processing.

\subsection{Privacy and Security Challenges}

The deployment of LLM and FM-based ECG systems introduces distinct privacy and security vulnerabilities across three architectural layers: the cloud, the edge, and the communication link between them. In the cloud, where large-scale models reside, critical attacks include adversarial prompt injection, jailbreaking, and model extraction. These can compromise safety alignment, leak sensitive biometric data, or enable unauthorized replication of model functionality. Li et al. \cite{li2024llm} show that preserving safety alignment during fine-tuning mitigates some of these cloud-side risks. On the edge, where SLMs, conventional DL models, or local AI agents operate, emerging agentic systems introduce additional attack surfaces \cite{he2024emerged}, such as local model poisoning or unauthorized access to raw ECG signals stored on the device. Data privacy is paramount because ECG signals constitute sensitive biometric information; studies \cite{jaff2024data} highlight risks associated with data collection practices in LLM ecosystems, which often extend to edge-cloud data pipelines. On the edge-cloud link, attacks such as membership inference threaten the confidentiality of training data transmitted during federated learning or model updates. To counter these threats, essential safeguards include differential privacy, federated learning, and secure multi-party computation \cite{yan2024protecting}. Privacy benchmarks such as LLM-PBE \cite{li2024llm} provide systematic evaluation for assessing deployment risks across all three layers.

%\subsection{Privacy and Security Challenges}
%The deployment of LLM and FM-based ECG systems introduces significant privacy and security concerns. Vulnerabilities such as adversarial prompt injection, jailbreaking, and model extraction attacks remain critical challenges. Li et al. \cite{li2024llm} show that preserving safety alignment during fine-tuning can mitigate some risks, while emerging agentic systems introduce additional attack surfaces \cite{he2024emerged}. Data privacy is particularly critical for ECG signals, which constitute sensitive biometric data. Studies \cite{jaff2024data} highlight risks associated with data collection practices in LLM ecosystems, while attacks such as membership inference threaten confidentiality. Techniques such as differential privacy, federated learning, and secure multi-party computation are therefore essential safeguards \cite{yan2024protecting}. Privacy benchmarks such as LLM-PBE \cite{li2024llm} provide systematic evaluation for assessing deployment risks.

\subsection{Discussion}

The literature indicates that hybrid compression strategies combining quantization, pruning, and knowledge distillation provide the most effective pathway for deploying ECG intelligence on edge devices, achieving compression ratios exceeding 100× with minimal accuracy loss. SLMs offer a promising balance between efficiency and capability, while TinyML enables ultra-low-power deployment for continuous monitoring. Federated learning complements these approaches by enabling privacy-preserving collaborative training.

Despite significant progress, challenges remain in achieving robust on-device training, handling hardware heterogeneity, and ensuring reliable performance across diverse populations and signal conditions. Some interesting research directions include ECG-specific small language models capable of multimodal reasoning on-device, adaptive edge-cloud hybrid systems, and exploration of emerging paradigms such as neuromorphic and in-memory computing. Rigorous clinical validation and standardized benchmarking frameworks is also essential to translate these advances into real-world cardiovascular healthcare systems.

In summary, enabling LLM and foundation model-driven ECG intelligence on the edge requires a holistic approach encompassing model compression, efficient architectures, federated learning, and hardware–software co-design, alongside strong privacy and security guarantees. These advances pave the way for scalable, real-time, and trustworthy cardiovascular monitoring systems operating directly on the edge.

%%%%%%%%%%%%%%%%%%%%%%%%%%%%%%%%%%%%%%%%%%%%%%%%%%%%%%%%%%%%

\section{Emerging Topics and Future Directions}

\begin{figure*}
    \centering
    \includegraphics[width=1\linewidth]{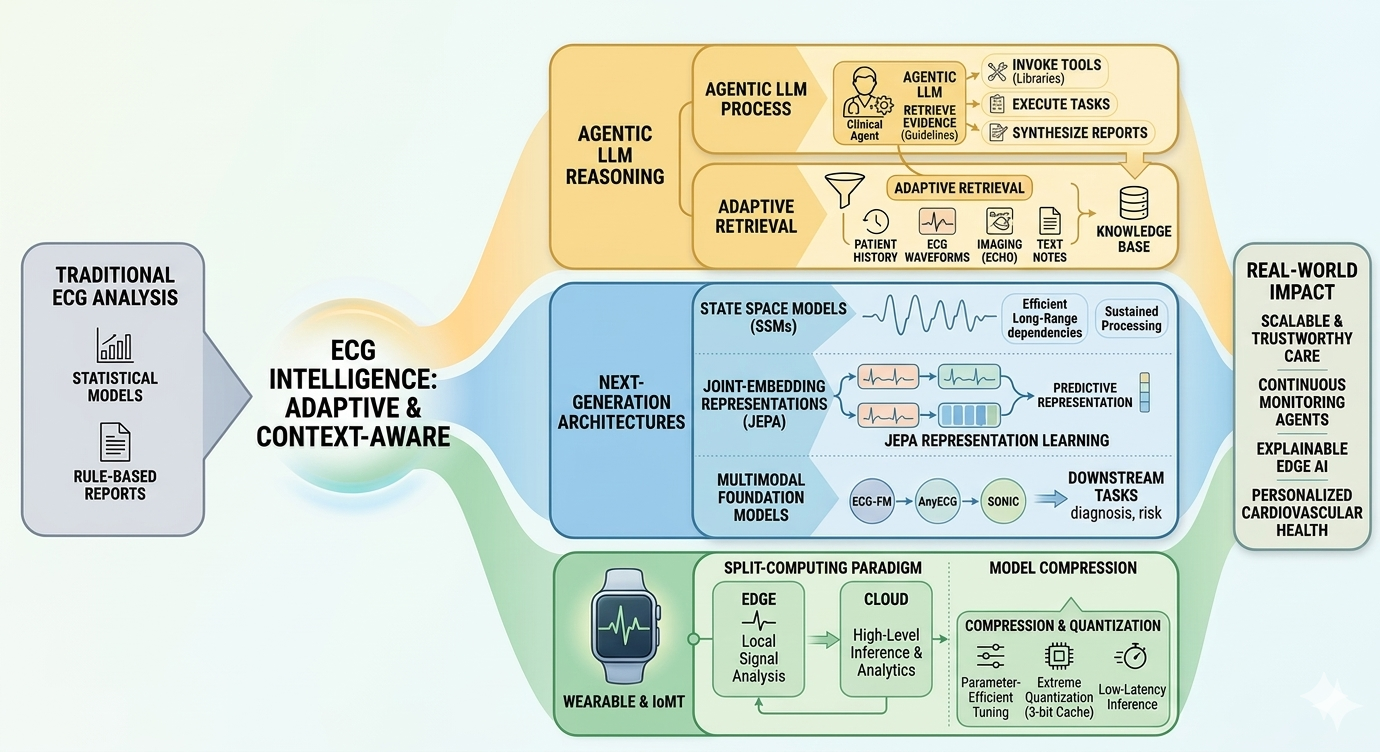}
    \caption{Outlook for edge-aware Agentic AI ECG intelligence.}
    \label{fig:vision}
\end{figure*}

The convergence of several recent advances—ECG foundation models acting as expert waveform interpreters, medical LLMs functioning as semantic reasoning engines, continued maturation of the edge AI ecosystem, and the emergence of autonomous AI agents—signals a paradigm shift in ECG intelligence: moving away from static interpretation pipelines toward adaptive, context-aware, and semi-autonomous clinical support systems.
Rather than functioning solely as predictive classifiers or report generators, next-generation agentic ECG AI systems are increasingly envisioned as orchestrators of multimodal data, clinical knowledge, and real-time physiological monitoring. Our synthesis identifies several transformative frontiers shaping this transition: the evolution from conventional retrieval-augmented generation (RAG) toward agentic reasoning workflows, the emergence of next-generation sequence architectures such as state space models and world models, the growing importance of edge-native intelligence enabled by SLMs, and exploratory directions spanning quantum acceleration, physical AI, and cognitive architectures. Collectively, these developments point toward a future in which ECG-driven cardiovascular intelligence becomes agentic, continuous, personalized, and deployment-aware.
Fig. \ref{fig:vision} provides a visual summary of the outlook for the AI-powered ECG intelligence.

\subsection{From RAG to Agentic Workflows}

While RAG has significantly improved factual and clinical relevance in LLM outputs by conditioning generation on curated medical knowledge bases, the field is rapidly advancing toward agentic paradigms in which models actively plan, reason, and utilize external tools. This evolution reflects a broader shift from passive information retrieval toward structured clinical task execution.

\subsubsection{Agentic RAG and Tool-Oriented Reasoning}

Contemporary RAG pipelines have progressed beyond naive document retrieval to hybrid architectures that integrate dense neural embeddings with sparse lexical matching for precise contextual grounding \cite{gao2023retrieval, shi2024ask}. Building on this foundation, agentic RAG frameworks position the LLM as a supervisory reasoning engine capable of decomposing complex clinical objectives into executable sub-tasks. For example, an ECG-focused clinical agent tasked with assessing arrhythmia risk may autonomously retrieve longitudinal patient history, invoke specialized signal analysis modules, and synthesize multimodal evidence into structured diagnostic reports \cite{tang2024electrocardiogramrag}.

Recent implementations such as LLM-powered physiological agents demonstrate improved robustness compared to static inference pipelines by maintaining persistent representations of patient states and dynamically selecting analytical tools for time-series interpretation. This paradigm reduces hallucination risk by grounding outputs in verifiable computational processes rather than purely probabilistic token generation \cite{alkhalaf2024applying}. Furthermore, emerging open agent frameworks and tool ecosystems are accelerating experimentation with modular ECG analysis agents capable of integrating signal processing libraries, clinical guideline retrieval engines, and decision support modules within unified reasoning loops.

\subsubsection{Dynamic and Multimodal Retrieval for Cardiovascular Reasoning}

Cardiovascular diagnosis inherently requires the integration of heterogeneous modalities, including one-dimensional ECG waveforms, imaging data such as echocardiography, and textual clinical documentation. Recent multimodal retrieval frameworks exemplified by systems such as ECG-LM and REALM illustrate how foundation models can bridge low-level signal representations with high-level semantic medical knowledge \cite{zhu2024realm}. 

Complementary advances in dynamic retrieval mechanisms, such as DRAGIN, allow LLMs to determine both \textit{when} and \textit{what} to retrieve during inference, thereby optimizing trade-offs between latency, computational cost, and contextual richness \cite{su2024dragin}. Such adaptive retrieval strategies are particularly relevant for continuous ECG monitoring scenarios, where timely interpretation must be balanced against resource constraints in wearable or edge environments.

\subsection{Next-Generation Architectures for ECG Analysis}

Transformer-based architectures have dominated recent advances in ECG modeling owing to their ability to capture long-range temporal dependencies and hierarchical waveform representations. However, their quadratic computational and memory scaling with sequence length presents fundamental challenges for continuous or long-duration cardiac monitoring, particularly in ambulatory and wearable settings where efficiency constraints are stringent.

\subsubsection{State Space Models}

To address these limitations, state space models (SSMs), including emerging architectures such as Mamba, are gaining traction as computationally efficient alternatives that achieve near-linear scaling with sequence length. These models enable sustained processing of extended physiological time-series while maintaining strong capacity to model complex temporal dynamics and rhythm variability \cite{hu2025adaptive}. Recent proposals such as MSECG and ECG-Mamba demonstrate that SSM-based approaches can effectively capture both long-term cardiac trends and transient anomalies with significantly reduced memory footprints compared to attention-centric models \cite{mckeen2024ecg}. Such efficiency advantages are particularly valuable for real-time arrhythmia surveillance, longitudinal monitoring, and edge-native inference scenarios.

\subsubsection{Joint-Embedding Predictive Architectures}
Beyond SSMs, Joint-Embedding Predictive Architecture (JEPA) paradigms are beginning to attract interest in biosignal modeling. By learning predictive representations through the alignment of contextual embeddings rather than explicit generative reconstruction, JEPA-style models (e.g., DSeq-JEPA and seq-JEPA) offer a promising pathway for self-supervised ECG learning that emphasizes semantic invariances across leads, patients, and acquisition conditions \cite{weimann2025self,kim2024learning}. This representation-centric approach may enable more robust generalization under domain shift and facilitate scalable pretraining on large unlabeled ECG corpora.

\subsubsection{Vision-Based and Multimodal Foundation Models}

In parallel, Vision Transformers and signal-to-image representation strategies continue to shape ECG research by transforming one-dimensional waveforms into spectrotemporal or spatial encodings suitable for large-scale visual foundation models. Frameworks such as SONIC illustrate the potential of ensemble vision architectures for stress detection and behavioral cardiology applications using wearable sensor streams \cite{phukan2024sonic}.

More broadly, cardiology-oriented foundation models such as ECG-FM and AnyECG are consolidating the shift toward large-scale self-supervised pretraining on millions of recordings, enabling strong performance across diverse downstream tasks including arrhythmia detection, cardiac function estimation, and prognostic risk modeling \cite{wang2024anyecg, li2024electrocardiogram}. By integrating multimodal inputs and transferable representations, these models represent an important step toward unified cardiovascular intelligence systems capable of generalizing across institutions, devices, and patient populations.

\subsection{Edge-Native Cardiovascular Intelligence and Ultra-Efficient Model Compression}

As ECG monitoring increasingly migrates toward wearable ecosystems and Internet of Medical Things (IoMT) platforms, reliance on centralized cloud inference introduces latency bottlenecks, intermittent connectivity risks, and heightened privacy exposure that may be incompatible with safety-critical cardiac applications. This has spurred interest in compact language and foundation models capable of edge-native deployment. Benchmarks such as HealthSLM-Bench show that instruction-tuned compact architectures—inspired by Phi-3 and TinyLlama—can perform competitive diagnostic reasoning on mobile hardware, enabling split-computing paradigms where local devices handle signal interpretation and anomaly detection while selectively offloading higher-level analytics to the cloud \cite{zhang2024tinyllama}.

Realizing this vision requires model compression techniques that reduce memory and computation without sacrificing clinical accuracy. Conventional methods (quantization-aware training, pruning, knowledge distillation) have made multimodal ECG pipelines feasible on constrained platforms. More recent advances address deeper bottlenecks, e.g., TurboQuant employs a two-stage zero-training pipeline combining PolarQuant transformations with residual quantized Johnson–Lindenstauss projections to reduce key–value caches to as low as 3 bits, achieving near-lossless long-context performance with substantial memory and latency reductions \cite{zandieh2025turboquant}.

Parallel efforts focus on maximizing intelligence density (useful intelligence per gigabyte) as a paradigm beyond parameter scaling. The open-source 1-bit Bonsai 8B model (a heavily quantized Qwen3) achieves over 10× the intelligence density of full-precision counterparts, fitting into 1.15 GB while delivering 14× smaller size, 8× faster inference, and 5× better energy efficiency on edge hardware \cite{prismml_bonsai_2026}. Similarly, BitNet b1.58 2B4T—a native 1‑bit LLM trained on 4 trillion tokens—matches the performance of full-precision 2B models across language, reasoning, and coding benchmarks, while dramatically reducing memory, energy, and latency \cite{ma2025bitnetb1582b4ttechnical}.

Collectively, these advances in compact reasoning models and ultra-efficient quantization enable agentic ECG AI to be embedded directly into wearable platforms. By doing on-device ECG analytics and basic reasoning, edge-native architectures support timely detection of critical cardiac events and facilitate privacy-aware, personalized care.

\subsection{Quantum Acceleration and Computational Frontiers}

As large-scale ECG datasets increasingly encompass continuous multi-lead recordings across diverse populations, computational scalability remains a central challenge. Quantum computing has been proposed as a potential long-term solution for accelerating model optimization and representation learning. For instance, rank-adaptive techniques such as QuanTA aim to enable efficient fine-tuning of large models in patient-specific contexts, potentially reducing adaptation complexity relative to classical approaches \cite{chen2024quanta}. 

In addition, quantum-inspired tensor compression methods such as CompactifAI explore the use of tensor networks to reduce the memory footprint of large neural models, achieving substantial parameter compression while preserving predictive performance \cite{tomut2024compactifai}. Although still largely exploratory, such approaches highlight the importance of interdisciplinary innovation in cardiovascular AI research.

\subsection{Toward Physical AI and Cognitive Cardiovascular Agents}

Beyond algorithmic improvements, emerging research in physical AI and embodied intelligence suggests that future cardiovascular monitoring systems may operate as persistent agents embedded within real-world environments. By integrating LLM-based reasoning with sensor fusion, control mechanisms, and causal world models of cardiac physiology, such systems could transition from passive monitoring toward proactive intervention and personalized health coaching. 

Current LLM-driven systems remain predominantly artificial narrow intelligence (ANI), excelling at pattern recognition but lacking robust causal reasoning required for complex differential diagnosis. To bridge this gap, researchers are exploring cognitive architectures that combine foundation models with structured memory, physiological simulators, and long-horizon planning capabilities \cite{shang2024ai}. Embodied deployment---ranging from intelligent wearables to assistive robotic platforms---will enable continuous interaction between AI agents and patients, redefining cardiovascular care.

\subsection{Ethical, Privacy, and Security Imperatives}

The integration of LLM-driven intelligence into IoMT ecosystems introduces significant privacy and security challenges. Model inversion and membership inference attacks, for example, can potentially expose sensitive patient information encoded within learned representations \cite{li2024llm}. 

Federated learning has therefore emerged as a promising paradigm for collaboratively training ECG foundation models without requiring centralized data aggregation. Modern privacy-preserving pipelines increasingly combine federated optimization with differential privacy mechanisms and homomorphic encryption to ensure that even intermediate training updates cannot be reverse-engineered \cite{yan2024protecting}. Additionally, trusted execution environments available in contemporary hardware platforms provide secure enclaves for processing sensitive physiological data, enabling hardware-level protection for real-time cardiac AI applications \cite{pati2024privacy}. 

Ensuring that future ECG intelligence systems are not only accurate and scalable but also secure, transparent, and ethically aligned will be essential for achieving sustained clinical adoption and public trust.

\section{Conclusion}

This survey has presented a unified perspective on the emerging paradigm of FM/LLM-enabled ECG intelligence for cardiovascular disease diagnosis, monitoring, and decision support. By tracing the evolution from expert-driven interpretation and conventional machine learning to deep learning and foundation-scale reasoning models, we have highlighted how contemporary medical LLMs and ECG foundation models are reshaping the way cardiac signals are analyzed, contextualized, and operationalized in clinical workflows. A key takeaway is that ECG analysis is no longer confined to isolated signal classification tasks; instead, it is increasingly framed as a multimodal reasoning problem that integrates physiological time-series data with clinical knowledge, patient history, and population-level insights.

The survey has further emphasized that methodological innovation must be accompanied by deployment awareness. Model compression strategies, edge-native architectures, and privacy-preserving learning frameworks are not peripheral engineering considerations but central enablers of real-world cardiovascular AI. As consumer wearable devices and continuous monitoring platforms generate unprecedented volumes of ECG data, scalable and trustworthy intelligence will depend on balancing model capacity with latency, energy efficiency, and data sovereignty constraints.

Looking ahead, emerging directions such as agentic AI workflows, multimodal cardiovascular foundation models, and adaptive edge-cloud ecosystems suggest a transition toward continuous, personalized cardiac intelligence. At the same time, ensuring interpretability, robustness, and ethical governance also remain critical for clinical acceptance and societal trust. By synthesizing advances across model design, ECG reasoning methodologies, and deployment strategies, this survey provides a roadmap to build next-generation cardiovascular AI systems that are not only accurate, but also context-aware, secure, and clinically impactful.

\footnotesize{
\bibliographystyle{IEEEtran}
\bibliography{references}}

\vfill\break

\end{document}